# Systèmes interactifs sensibles aux émotions : architecture logicielle

*Alexis Clay*
*LIPSI-ESTIA / IBRLab-NTHU*
*Technopole Izarbel, 64210 Bidart*
a.clay@estia.fr

**Résumé**
Nous présentons une architecture logicielle pour des systèmes interactifs qui intègrent l'émotion de l'utilisateur. Cette dernière peut être utilisée au sein d'un système interactif à plusieurs fins : changement des fonctionnalités d'un système interactif comme dans un système d'enseignement par simulateur [1], de conseils médicaux [2].
Nous verrons que dans notre domaine d'application, le spectacle de ballet, l'émotion est un concept manipulé explicitement par le système interactif qui la rend perceptible selon les différentes modalités en sortie.
Notre solution architecturale pour développer des systèmes interactifs sensibles aux émotions repose sur l'ajout d'une nouvelle branche de capture-analyse-interprétation au modèle architectural PAC-Amodeus. Nous montrons les différents flots d'information au sein de cette architecture selon l'impact de l'émotion capturée sur le système interactif. Nous illustrons enfin notre modèle architectural en considérant notre application de spectacle de ballet. L'objectif est d'augmenter en temps réel une scène de spectacle pour sublimer l'expérience artistique vécue par le public. Ainsi le système interactif capture l'émotion produite par un danseur pendant son jeu. Le système choisit alors un objet virtuel dynamique à afficher sur la scène et adapte ses paramètres en fonction de l'émotion jouée et de son intensité en temps réel, contribuant ainsi à l'expérience artistique.

**Introduction**

L'émotion est encore un domaine peu étudié en Interaction Homme-Machine (IHM). Pourtant, l'homme est une créature profondément émotionnelle, qui se repose sur l'affect bien plus que sur la logique ou sur la raison. Chaque décision que nous prenons fait intervenir l'affect : des expériences menées sur des malades ne pouvant ressentir certaines émotions montrent qu'ils sont incapables de prendre la moindre décision [3], [4]. L'émotion joue également un rôle dans les mécanismes de mémorisation puisque les événements associés à une émotion sont bien mieux mémorisés que les autres [4]. Cette dimension humaine a été totalement ignorée jusqu'à il y a quelques années en IHM. L'importance de l'affect dans notre façon de penser, de raisonner, et d'interagir avec tout ce qui nous entoure justifie les études sur les systèmes réactifs à l'état affectif de l'utilisateur.

Nos travaux concernent les systèmes interactifs qui sont réactifs à l'état affectif de l'utilisateur : nous visons la capture-analyse-interprétation de l'émotion de l'utilisateur, celle-ci ayant ensuite un impact sur le système interactif. Dans le domaine lié aux émotions (« affective computing » [3]) d'autres travaux concernent par exemple la synthèse d'émotion comme dans les systèmes de dialogue naturel avec un avatar [2].

Cet article est consacré aux aspects ingénierie logicielle des systèmes interactifs sensibles aux émotions de l'utilisateur et nous présentons une extension du modèle PAC-Amodeus dédiée aux systèmes interactifs multimodaux. Nous montrons les différents flots d'information au sein de cette architecture selon l'impact de l'émotion capturée sur le système interactif. Nous illustrons enfin notre modèle architectural en considérant notre application de spectacle de ballet. L'objectif est d'augmenter en temps réel une scène de spectacle pour sublimer l'expérience artistique vécue par le public. Ainsi le système interactif capture l'émotion produite par un danseur pendant son jeu. Le système choisit alors un objet virtuel dynamique à afficher sur la scène et adapte ses paramètres en fonction de l'émotion jouée et de son intensité en temps réel, contribuant ainsi à l'expérience artistique.

L'article est organisé comme suit : nous définissons d'abord les notions de base qui caractérisent une émotion puis nous exposons notre extension du modèle architectural PAC-Amodeus et détaillons les différents flots d'information au sein de notre modèle. Nous illustrons enfin notre modèle en considérant notre système de spectacles de ballet.

**Les bases de l'émotion**

Il existe de nombreuses définitions du terme "émotion" [5][6][7]. Cependant, la plupart se rejoignent sur un point : une émotion est un processus physio-psychologique de forte intensité en réaction à un stimulus. Le réseau d'excellence européen HUMAINE [8], dont le but est d'étudier les aspects liés aux émotions dans les systèmes informatiques, fournit une définition structurée, se basant sur la notion d'état affectif. Cette notion regroupe tout ce que nous pouvons ressentir et a été divisée en cinq catégories : les émotions, les humeurs, les positions interpersonnelles, les préférences/attitudes, et les dispositions affectives (Figure 1, tirée de [8]).

| Types of Affect | Intensity | Duration | Synchronization | Event focus | Appraisal elicitation | Rapidity of change | Behavior impact |
|---|---|---|---|---|---|---|---|
| **Emotions:** *angry, sad, joyful, fearful, ashamed, proud, elated, desperate* | ● | · | ● | ● | ● | ● | ● |
| **Moods:** *cheerful, gloomy, irritable, listless, depressed, buoyant* | ● | ● | · | · | · | ● | · |
| **Interpersonal stances:** *distant, cold, warm, supportive, contemptuous* | ● | ● | · | ● | · | ● | ● |
| **Preferences/Attitudes:** *liking, loving, hating, valuing, desiring* | ● | ● | · | · | · | · | ● |
| **Affect dispositions:** *nervous, anxious, reckless, morose, hostile* | · | ● | · | · | · | · | · |

**Figure 1: Espace de caractérisation des états affectifs. Tableau issu de [8]**

Selon cette définition, une émotion est caractéristique sur tous les points :

- Une forte intensité ;
- Une émotion est une réaction de courte durée ;
- Une forte synchronisation : tout le corps réagit à l'unisson ;
- une émotion est directement liée à un évènement déclencheur ;
- Une émotion est cependant soumise au traitement cognitif de l'évènement déclencheur ;
- Une émotion peut changer très rapidement ;
- Son impact sur le comportement est important.

Darwin, dans sa théorie sur l'évolution, souligne que les émotions sont une réponse à l'environnement apparues de la même façon que bien d'autres phénomènes, par sélection naturelle. Une émotion induit des réactions physiologiques et psychologiques nous permettant de mieux répondre à l'environnement. Paul Ekman, fervent partisan de la théorie darwiniste, a extrait six émotions « basiques » en se basant sur six critères [4] :

- Elles doivent être déclenchées par des stimuli universels, soit communs à tous les membres de l'espèce ;
- Elles doivent apparaître spontanément ;
- La réaction doit être rapide à apparaître et à disparaître ;
- Le traitement cognitif du stimulus doit être automatique ;
- Elle doit déclencher des pensées ou sensations spécifiques ;
- Elle doit être présente chez d'autres primates que chez l'humain.

Les six émotions extraites à travers ce filtre sont la colère, la tristesse, la peur, la joie, la surprise et le dégoût. Les contraintes imposées par Ekman pour son choix rejoignent la définition d'une émotion par le réseau d'excellence HUMAINE. Contrairement aux émotions de premier ordre – les émotions basiques - dont on peut voir les manifestations chez le nourrisson, les émotions de second ordre sont celles qui sont passées par le filtre de l'expérience et de l'apprentissage social. Les six émotions basiques d'Ekman, largement acceptées par la communauté des psychologues, fournit un premier ensemble, discret, d'émotions sur lequel se baser.

Une deuxième façon de catégoriser l'émotion s'appuie sur un espace continu d'émotions. Cet espace multi-dimensionnel a été ramené à deux dimensions par Plutchik et par Whissel [3]. Plutchik définit ses deux axes comme la valence de l'émotion (déplaisant-plaisant) et l'excitation de l'émotion (faible-forte). Les émotions basiques se retrouvent a la périphérie d'un cercle centre a l'origine sur la représentation d'un tel espace, avec l'ensemble des émotions contenu a l'intérieur de ce cercle.

Nous retenons donc deux façons de catégoriser l'émotion. La première est l'approche discrète, ou l'on définit chaque émotion, la deuxième est l'approche continue. La définition adoptée de l'émotion a un impact direct sur la partie du code qui est responsable de la capture-analyse-interprétation de l'émotion. Dans le paragraphe suivant nous adoptons une approche globale et nous visons à localiser la partie de code dédiée à l'identification de l'émotion au sein d'une architecture logicielle de système interactif et à étudier comment l'émotion captée et interprétée est exploitée au sein du système interactif.

**Modèle d'architecture pour des systèmes interactifs sensibles aux émotions**
Tout en étant indépendant du type de catégorisation des émotions adopté – ensemble discret ou continu - nous proposons un modèle d'architecture logicielle pour systèmes interactifs sensibles aux émotions. Notre solution consiste à étendre le modèle PAC-Amodeus. Dans le paragraphe suivant nous rappelons brièvement les composants du modèle PAC-Amodeus et leurs rôles. Nous exposons ensuite notre extension puis nous étudions les différents flots d'information.

**Le Modèle PAC-Amodeus**
Le modèle PAC-Amodeus se fonde sur le modèle en 5 composants Arch [9] (Figure 2). Cette modélisation en cinq composants indépendants permet une grande modifiabilité du système. Le contrôleur de dialogue (CD) est la clé de voûte du système et sert de lien entre les branches fonctionnelles et de présentation.
La branche fonctionnelle est constituée d'un Noyau Fonctionnel (NF) et d'une Interface au Noyau Fonctionnel (INF). Le noyau fonctionnel regroupe les fonctions abstraites relatives au concept. Ces fonctions sont totalement orientées informatiques et indépendantes de la représentation du concept par l'utilisateur. L'interface au noyau fonctionnel sert de lien entre le noyau fonctionnel et le contrôleur de dialogue. Son rôle est de traduire les concepts purement informatiques du noyau fonctionnel en concepts plus orientés tâches utilisateur.
La branche de présentation est constituée d'un Composant d'Interaction Physique (CIL) et d'un composant d'Interaction Logique (CIL). Le premier est l'interface au plus bas niveau et dépend du système utilisé. Le deuxième traduit ces informations pour les envoyer au contrôleur de dialogue.

Le modèle PAC-Amodeus affine le modèle Arch en définissant le contrôleur de dialogue comme une hiérarchie d'agents PAC. Un agent PAC est constitué de trois facettes : Présentation, Abstraction, Dialogue [Figure 2]. Les facettes Présentation et Abstraction doivent passer par la facette Dialogue pour communiquer. Le modèle PAC-Amodeus est complètement présenté dans [9].

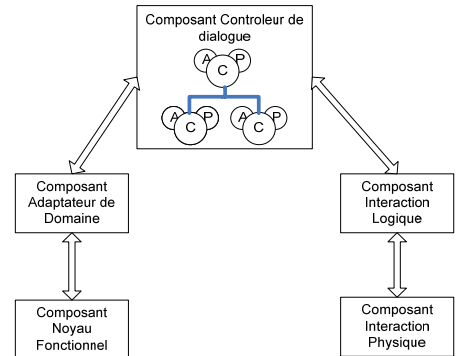

**Figure 2 : Le modèle PAC-Amodeus**

**Extension du modèle PAC-Amodeus**
Par analogie avec l'extension proposée pour la capture de contexte [10] nous identifions trois nouveaux composants qui correspondent aux étapes classiques de capture-analyse-interprétation. Ces trois composants constituent une nouvelle branche au sein du modèle PAC-Amodeus en appliquant le mécanisme de « branching » hérité du modèle ARCH. Dans la Figure 3, nous montrons ces trois composants liés au composant Contrôleur de Dialogue du modèle PAC-Amodeus.

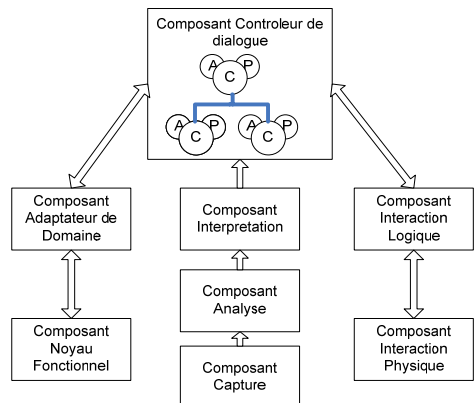

**Figure 3 : Modèle PAC-Amodeus étendu**

Les deux composants Capture Analyse sont dépendants du système de capture utilisé. La capture peut être effectuée selon plusieurs modalités : faisant l'analogie avec les modalités d'interaction [9] nous retrouvons les deux composants Capture et Analyse qui correspondent aux deux composants Interaction Physique (dispositif) et Interaction Logique (langage d'interaction). Les données brutes peuvent être captées par la voix, le mouvement et la position du corps, par le visage, ou par les signaux physiologiques. Ces données sont ensuite transmises au Composant Analyse qui en fait l'abstraction en information utile à l'identification de l'émotion. Ainsi, pour le visage, on retiendra les mouvements du visage en se basant sur le Facial Action Coding System (FACS) ou les Facial Animation Points de M-PEG4 [13]. On peut également déterminer les éléments à mesurer via un apprentissage de l'ordinateur [14].

De l'analogie avec les modalités d'interaction nous identifions deux retombées intéressantes pour l'architecture :

Tout d'abord, le système de capture de l'émotion peut reposer sur des modalités d'interaction. Dans ce cas, la branche Emotion et la branche Interaction peuvent partager des composants. Par exemple, dans le cas d'un système interactif qui autorise des commandes vocales en entrée et qui est sensible aux émotions issues de l'analyse de la voix, la feuille de la branche Emotion et Interaction est commune (Dispositif microphone). L'analyse faite est ensuite différente, le composant Interaction Logique déduisant la commande vocale tandis que le composant Analyse extrait des caractéristiques du signal vocal utiles à l'identification d'une émotion (Figure 4).

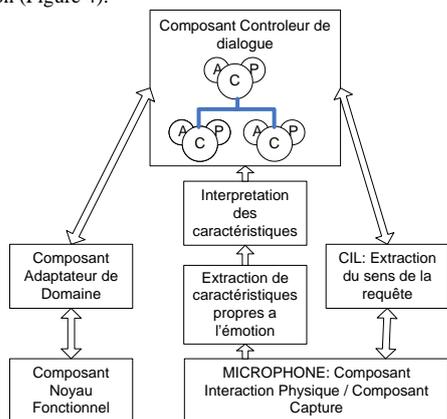

**Figure 4 : Fusion des Composants Capture / Interaction Physique**

De plus comme pour l'interaction multimodale [11] nous identifions un moteur de fusion/fission pour les données analysées. Plusieurs sources d'information peuvent être combinées de façon redondante ou complémentaire (propriétés CARE de la multimodalité [9]) au sein du composant Interprétation. Le modèle présenté figure 5 capte l'émotion grâce à la voix, aux mouvements du visage et à la gestuelle. Chaque modalité est capturée par un composant de capture adéquat ; les données sont ensuite traitées par les composants Analyse de chaque branche puis interprétées par les composants Interprétation.

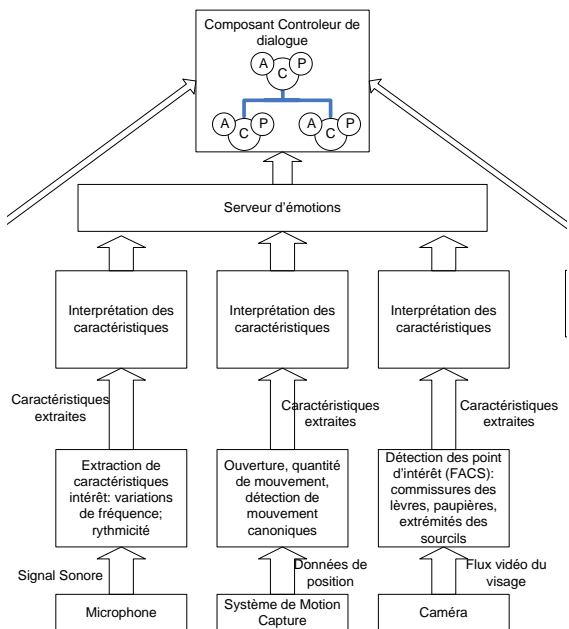

**Figure 5 : Multi-modalité de capture des émotions**

***Connexion de la nouvelle branche Emotion aux composants PAC-Amodeus***

A la figure 3, nous avons présenté le cas où la nouvelle branche est connectée au Contrôleur de Dialogue. Selon l'impact de l'émotion sur le système interactif, le flot de données est différent. Nous identifions trois cas :

- Cas 1 : L'émotion peut être explicitement manipulée par la branche fonctionnelle du modèle PAC-Amodeus. Par exemple dans notre application de spectacle de ballet, l'émotion captée est présentée éventuellement d'une façon accentuée. Aussi l'émotion est un concept du domaine qui sera manipulé par le Noyau Fonctionnel (Figure 6).

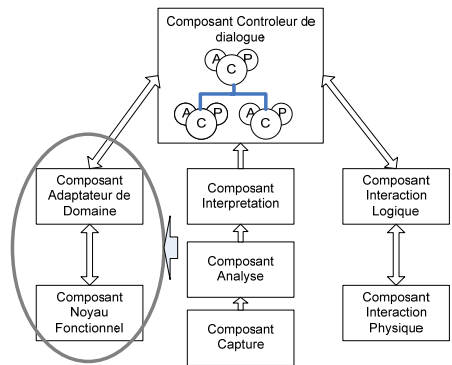

**Figure 6 : émotion manipulée par la branche fonctionnelle**

- Cas 2 : L'émotion peut avoir un impact sur l'enchaînement des tâches du Contrôleur de Dialogue. Dans ce cas, la branche dédiée à l'émotion est connectée au Contrôleur de Dialogue comme le montre la figure 3. Par exemple dans un système interactif d'enseignement assisté par ordinateur, la reconnaissance d'une émotion de tristesse, par exemple basée sur l'analyse du visage de l'apprenant, peut engendrer un nouveau dialogue d'aide sur l'exercice en cours. L'émotion détectée a alors un impact sur le Contrôleur de Dialogue du système interactif.

- Cas 3 : L'émotion détectée peut enfin avoir un impact sur la branche Interaction (figure 7). Par exemple une émotion détectée peut impliquer le changement de modalités d'interaction en sortie. De plus l'émotion peut aussi influencer les mécanismes liés à l'interaction en entrée comme par exemple rendre plus robuste la reconnaissance de commandes vocales en prenant en compte l'émotion détectée.

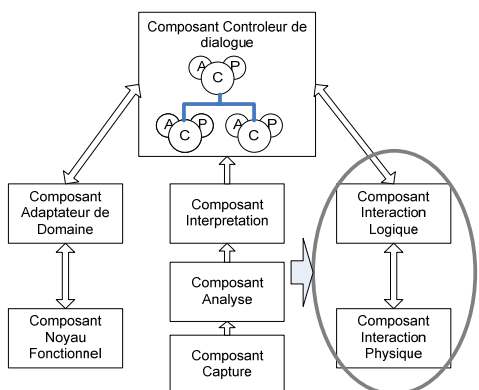

**Figure 7 : Emotion manipulée par la branche Interaction**

**Exemple illustratif : spectacle de ballets**

Dans le cadre de notre projet nous développons un système de spectacle permettant d'augmenter une scène de spectacle selon les émotions que le danseur cherche a transmettre au public. Pour cela, nous nous appuyons sur un système de capture du mouvement du type « combinaison de Capture du Mouvement »[12]. Quel que soit le système utilisé il existe des formats standards de sortie des données du système de capture, tel le format de fichier BVH. Le système de capture du mouvement est donc le composant Capture dans notre branche Emotion. Les données envoyées au composant Analyse sont les données brutes de capture dans un format standard quelconque.

Le module d'analyse prend donc en entrée un flot de données des positions de chaque joint du corps au cours du temps. Il traite alors ces données et en extrait des caractéristiques utilisables par la suite. Ainsi la quantité de mouvement produite par l'utilisateur, son expansion, et certains mouvements canoniques comme la position des bras (ouverts/fermes), l'orientation du tronc (gauche/droite) et sa position (cambre-rond)[15] sont calculés et mesurés : le rôle du composant analyse est de transformer les données brutes en éléments plus abstraits. Toutes ces mesures effectuées sont ensuite envoyées au module d'interprétation. Grace à elles, celui-ci détermine, selon ses propres règles, l'émotion exprimée *via* la modalité mesurée dans l'expression de l'utilisateur. Par exemple, le composant Analyse annonce au composant Interprétation que le corps est étiré, dirigé vers le haut, bras ouverts, dans un mouvement rapide, direct, dirigé vers l'avant et exécuté avec force. Le composant Interprétation, au vu de ces données, détermine l'émotion comme étant la joie [15]. Un mouvement identique mais dirigé vers l'arrière avec une fermeture des bras sera identifié comme la surprise.

Dans notre cas, l'émotion déterminée par la branche Emotion est directement prise en compte par le noyau fonctionnel : en effet ici l'émotion est au cœur du système puisqu'elle va définir et moduler ce qui sera affiché par la suite. Le noyau fonctionnel manipule directement ce concept d'émotion et envoie ses traitements au contrôleur de dialogue. Ce dernier fait ensuite appel à la branche Interaction afin de définir ce qui va être affiché afin de rendre ce que le public doit percevoir. Par exemple si le danseur exprime la colère : cette émotion est envoyée via l'INF au NF, qui peut décider de transmettre la même émotion ou au contraire de transmettre la peur dans le reste de la scène afin de créer un contraste. Ceci est ensuite traité par le CD, qui demande à la branche Interaction d'afficher les éléments en correspondance avec la décision du NF. Ainsi le public pourra observer des flammes, des auras rouges pour magnifier la colère, ou au contraire des teintes sombres, bleutées, afin de contraster avec l'émotion que le danseur transmet.

Un intérêt de cette architecture est sa modifiabilité. Ainsi, les composants Interprétation et Analyse sont indépendants du composant Capture : dans le cadre de capture du mouvement il suffit de changer le module de capture pour utiliser une autre technologie de motion capture. Les modules d'analyse et d'interprétation sont également indépendants : Remplacer le module d'interprétation permet donner une toute autre signification à des patterns de mouvement ou à des mouvements canoniques, sans changer les autres modules de la branche.

**Conclusion**

Nous avons vu dans cet article, une extension au modèle PAC-Amodeus permettant de modéliser un système adaptatif prenant en compte l'émotion. Cette extension prend la forme d'une nouvelle branche dans la lignée du mécanisme de branching hérité du modèle Arch et comporte trois composants : les composants Capture, Analyse, et Interprétation. Cette décomposition en éléments indépendants permet une grande modifiabilité du système : tant que les entrées/sorties des composants restent communes, le cœur du composant peut être changé à loisir.

Nous sommes actuellement en cours de développement de notre système de spectacle de ballets que nous testerons ensuite avec les ballets de Biarritz. L'architecture proposée sera alors mise à l'épreuve en testant sa modifiabilité et son extensibilité en considérant différentes méthodes de capture d'émotion en situation et plusieurs modalités d'interaction en sortie pour rendre perceptibles les émotions identifiées en temps réel lors du spectacle.

En perspectives à ces travaux, nous souhaitons aussi étudier l'impact des composants du modèle architectural sur la branche Emotion en particulier pour l'étape d'interprétation. En effet dans nos travaux, nous n'avons pour l'instant étudié que l'impact de la branche Emotion sur les autres composants du modèle.